\documentclass[letterpaper, 10 pt, conference]{arxiv}  

\IEEEoverridecommandlockouts                              

\usepackage{graphics} 
\usepackage{epsfig} 
\usepackage{amsmath} 
\usepackage{amssymb}  
\usepackage{color}
\usepackage[table,xcdraw,dvipsnames]{xcolor}
\usepackage[hidelinks]{hyperref}

\def\publishedurl{\centering \textbf{{\color{red}{The published version can be accessed from}} \color{RedViolet}{\url{https://doi.org/10.1109/ICRA.2019.8794286}}}}

\title{\LARGE \bf
Automatic Real-time Anomaly Detection for Autonomous Aerial Vehicles
}

\author{Azarakhsh Keipour$^{ 1}$, Mohammadreza Mousaei$^{ 2}$ and Sebastian Scherer$^{ 3}$
\thanks{* This work was supported through NASA Grant Number NNX17CL06C.}
\thanks{$^{1,2,3}$ Robotics Institute, Carnegie Mellon University, Pittsburgh, PA
        {\tt\small [keipour, mmousaei, basti]@cmu.edu}}%
}

\begin{document}

\maketitle
\thispagestyle{empty}
\pagestyle{empty}

\begin{abstract}

The recent increase in the use of aerial vehicles raises concerns about the safety and reliability of autonomous operations. There is a growing need for methods to monitor the status of these aircraft and report any faults and anomalies to the safety pilot or to the autopilot to deal with the emergency situation. In this paper, we present a real-time approach using the Recursive Least Squares method to detect anomalies in the behavior of an aircraft. The method models the relationship between correlated input-output pairs online and uses the model to detect the anomalies. The result is an easy-to-deploy anomaly detection method that does not assume a specific aircraft model and can detect many types of faults and anomalies in a wide range of autonomous aircraft. The experiments on this method show a precision of $88.23\%$, recall of $88.23\%$ and $86.36\%$ accuracy for over 22 flight tests. The other contribution is providing a new fault detection open dataset for autonomous aircraft, which contains complete data and the ground truth for 22 fixed-wing flights with eight different types of mid-flight actuator failures to help future fault detection research for aircraft.

\end{abstract}


\section{INTRODUCTION} \label{sec:introduction}

Technology for Autonomous Aerial Vehicles (AAVs) has tremendously advanced in recent decades to include a wide range of applications from security and traffic surveillance to the management of natural risks, environment exploration, agriculture, recreation, delivery, and enhancing flight experience of the hobby and commercial pilots. Despite the increasing use of AAVs, they do not exhibit the level of performance and reliability required to complete most missions autonomously, and there is still a general concern for their safety and reliability. 

To address safety concerns, the Federal Aviation Administration (FAA) has suggested a series of regulations (e.g., Title 14 Code of Federal Regulations) about AAV safety requirements \cite{faa}. A standout amongst the most vital concerns for reliability is the behavior of the system during a breakdown, which raises the need for AAVs to have the capacity to detect faults in the system and react accordingly. 

Larger aircraft usually devise redundant hardware to address the safety and reliability concern, which is a reliable approach, but is more expensive, adds weight and occupies more space. However, for smaller aircraft, including smaller AAVs, the hardware redundancy is generally not possible due to space and load constraints. To provide the necessary reliability to these aircraft, a Fault Detection, Isolation and Recovery (FDIR) is required. Fault Diagnosis is a fundamental piece of FDIR techniques which can be divided into three sections: Fault Detection, Fault Isolation and Fault Estimation. Fault Detection is to recognize if a problem has happened; Fault Isolation is to decide the area in which the fault has occurred; Fault Estimation is to find the type of fault and its impacts.

\begin{figure}[t]
      \centering
      \includegraphics[width=0.48\textwidth]{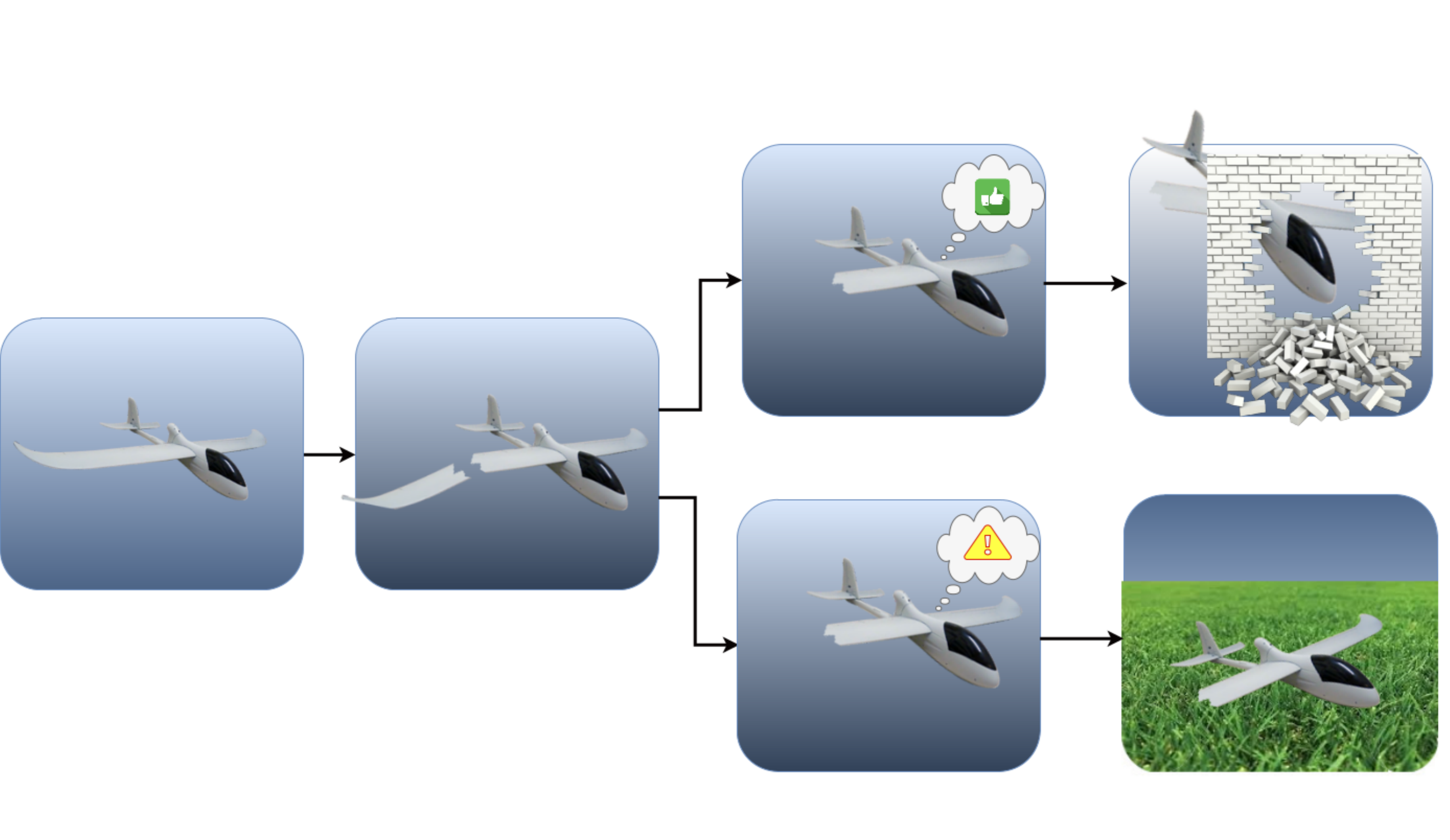}
      \vspace{-0.7cm}
      \caption{An overview of the problem: in the occasion of an anomaly or a fault occurring in the AAV system, not having knowledge about the fault can be a great threat for the AAV safety (upper right image). However, AAV may perform emergency reaction if the fault is immediately detected.}
      \label{fig:story}
\end{figure}

Our contributions include: 1) Providing an easy-to-deploy method for detecting anomalies in the behavior of aircraft due to sudden actuator, sensor or component faults. Our Anomaly Detection method is simple enough to be easily implemented and integrated into a new type of aircraft and does not assume a specific aircraft model. It is computationally low-cost and can detect many types of faults and failures in a wide range of autonomous aircraft (e.g., large and small single-rotor aircraft, multirotors, fixed-wings). 2) We provide a new open dataset for autonomous aircraft fault detection. It consists of complete data from 22 flights containing eight types of failures of 4 different actuators on a real fixed-wing UAV. The dataset provides the ground truth for the failures (exact time and the type of failures) to help future research in the fault detection area.

\section{BACKGROUND} \label{sec:background}

In past decades, many new Fault Detection methods were introduced in response to the concerns for the safety of aircraft operations. Many of these methods target only specific types of aircraft: Kuric et al. \cite{8014173} and Han et al. \cite{954410017691794} present approaches for use in multirotors; Qi et al. \cite{6564801, AReviewonFault} review different methods for Fault Diagnosis in helicopters; Melody et al. \cite{1-s2.0-S0967066100000460-main} and Ansari et al. \cite{7526603} describe methods for icing detection and sensor fault detection in fixed-wings, respectively.

A fault is defined as any undesired deviation of one or more parameters of a system from the standard conditions \cite{VANSCHRICK1997959}. In an aircraft, faults can be classified as actuator faults, sensor faults and plant (or component or parameter) faults \cite{6564801}. Actuator faults include partial or total loss of an actuator's control, which can result in a constant output (e.g., a stuck rudder or an engine failure), change in the actuator gains (e.g., partial loss of engine power), or drift in output values (e.g., change in the trim of the elevator). Sensor faults represent wrong measurement readings by the sensors, which can result in total faults (e.g., a random output from a faulty sensor), bias faults (e.g., bias in gyroscope reading), gain faults (e.g., in uncalibrated range sensor) and outlier faults (e.g., jumps in GPS reading). Plant faults include problems that change the dynamic properties of the system (e.g., a damaged wing) and the complete loss of communication between the controller and a component \cite{7069265}.

Many of the fault detection methods are developed only to detect a specific set of faults. For example, Melody et al. \cite{1-s2.0-S0967066100000460-main} and Cristofaro et al. \cite{icing1} present methods for icing detection in fixed-wing aircraft, Kuric et al. \cite{8014173} describe a method for the detection of propulsion system faults in octorotors, and Ducard \cite{10.10072F978-90-481-9707-143} focuses on actuator fault detection. 

Various types of approaches exist for the task of fault detection \cite{AReviewonFault, Khalastchi:2018:FDD:3177787.3146389}. Analytical and model-based approaches devise mathematical models. In many cases, the dependency on the accurate prior modeling of machines with such complexity as aircraft makes it hard to apply these methods for other types of aircraft or faults. 
Another class of methods is signal processing-based approaches. These methods avoid the need for the aircraft model, and they determine the faults through analysis of the signals available in the aircraft. Knowledge-based approaches use information related to the faults and devise the knowledge of experts to detect different types of faults without the need to model the aircraft accurately. However, transferring the developed knowledge-based methods from an application to another application or extending them to cover different types of faults is very challenging. 

Many of the available methods are a combination of two or more classes. 
The authors in Melody et al. \cite{1-s2.0-S0967066100000460-main} use three different methods for parameter estimation to monitor changes in dynamics model for icing detection on fixed-wing aircraft: Batch Least Squares method, Extended Kalman Filter and $H^{\infty}$ algorithm. The tests in simulation show that only $H^\infty$ works for the purpose, which relies on a good model and assumes that state derivative information is available. 
The algorithm presented by Birnbaum et al. \cite{UAVSecurity0167FrDTT2-04} compares the time spent on different segments of the flight plan with the actual flight times to detect unplanned flight deviations indicative of a cyber attack, sensor spoofing, or structural failure. The method avoids the need for a good model but depends on the availability of an accurate flight plan that already respects the model constraints.

Recursive Least Squares (RLS) is a standard signal processing-based technique for filtering and parameter estimation. 
The approach proposed by Kuric et al. \cite{8014173} diagnoses propulsion system faults in multirotors using RLS for controller parameter estimation and tests the method in simulation on a triple motor fault scenario.
The method by Han et al. \cite{954410017691794} uses RLS for quadrotor actuator fault detection and devises a parity space approach to estimate the forgetting factor for RLS. 
The method proposed by Birnbaum et al. \cite{birnbaum2015} applies RLS to the given model for UAV and performs estimation and tracking of the controller parameters. It processes the data in batches of size 500 samples and compares the parameters of different batches to detect discrepancies indicating anomalies. 

Khalastchi et al. \cite{p115-khalastchi} perform anomaly detection for different types of unmanned vehicles by first learning correlated input-output pairs of the system and then calculating Mahalanobis distance between data batches resulted from a sliding window on each stream of correlated input-output pairs. 

Other notable methods include Neural Network-based methods used for fault diagnosis and recovery \cite{10873863, neural-observer}, a Fuzzy Inference System (FIS) decision system combined with Particle Filter used for GPS fault detection on a hexarotor \cite{sensors-17-02243}, and methods utilizing Unscented Kalman Filter for sensor and actuator fault detection \cite{4252507, 4603358, IZADI20116343}.  

The method presented in this paper is a new real-time signal processing-based approach using the RLS method to detect anomalies in the behavior of an aircraft. The proposed method shares some basic ideas with \cite{birnbaum2015}, but in contrast to the methods modeling the entire aircraft, we only model the relationship between arbitrary correlated input-output signal pairs and rather than building the model prior to the flight, we estimate the model online. The proposed method is independent of the type of the aircraft, is not doing any batch processing (making it real-time) and does not assume any specific fault models. It can detect a wide range of anomalies in the behavior of various types of vehicles. We show the performance of the method on real field tests consisting of different fixed-wing actuator and engine faults.

\section{PROBLEM DEFINITION} \label{sec:problem}

The integration of a fast and reliable fault detection method can make a difference between a crash and a safe emergency maneuver for landing (See Fig~\ref{fig:story}). There is a need to develop a Fault Detection System that:
\begin{itemize}
\renewcommand\labelitemi{--}
\item Has minimal or no dependency on the type and model of the aircraft,
\item Can detect a wide range of anomaly types,
\item Does not rely on external information (e.g., flight path),
\item Works in real-time with a low computation cost,
\item Is simple to implement and can easily be integrated into a new aircraft,
\item Shows high performance in practice and not just in simulation.
\end{itemize}

Almost all the available methods for fault detection in AAVs depend on the model of the aircraft and are not able to easily be ported between two aircraft types. Some methods reduce this dependency by learning the dynamics of the aircraft from the past flights, but these methods rely on the assumption of a constant model, which is not practical in many applications where the dynamics can change from a flight to another (e.g., package delivery). Many of the methods are using characteristics of one or more specific sensors or modules to detect anomalies and faults in their outputs and behavior. This is also limiting the approach to detection of just specific fault types in the system. Some other methods heavily depend on external information such as accurate flight plans available to them. However, this is very impractical in many autonomous operations where trajectories change in flight to avoid collisions or to fly in unknown or partially-known environments. Finally, most of the proposed methods so far have never been tested in the field and have only shown results from software and Hardware-in-the-Loop simulations.

Our work aims to provide a solution to the fault detection problem that has all the mentioned characteristics needed for a Fault Detection system. The next section describes our method, its underlying theory and the assumptions for its reliable performance. 

\begin{figure}[!t]
\centering
    \includegraphics[width=0.47\textwidth, height=0.4\textwidth]{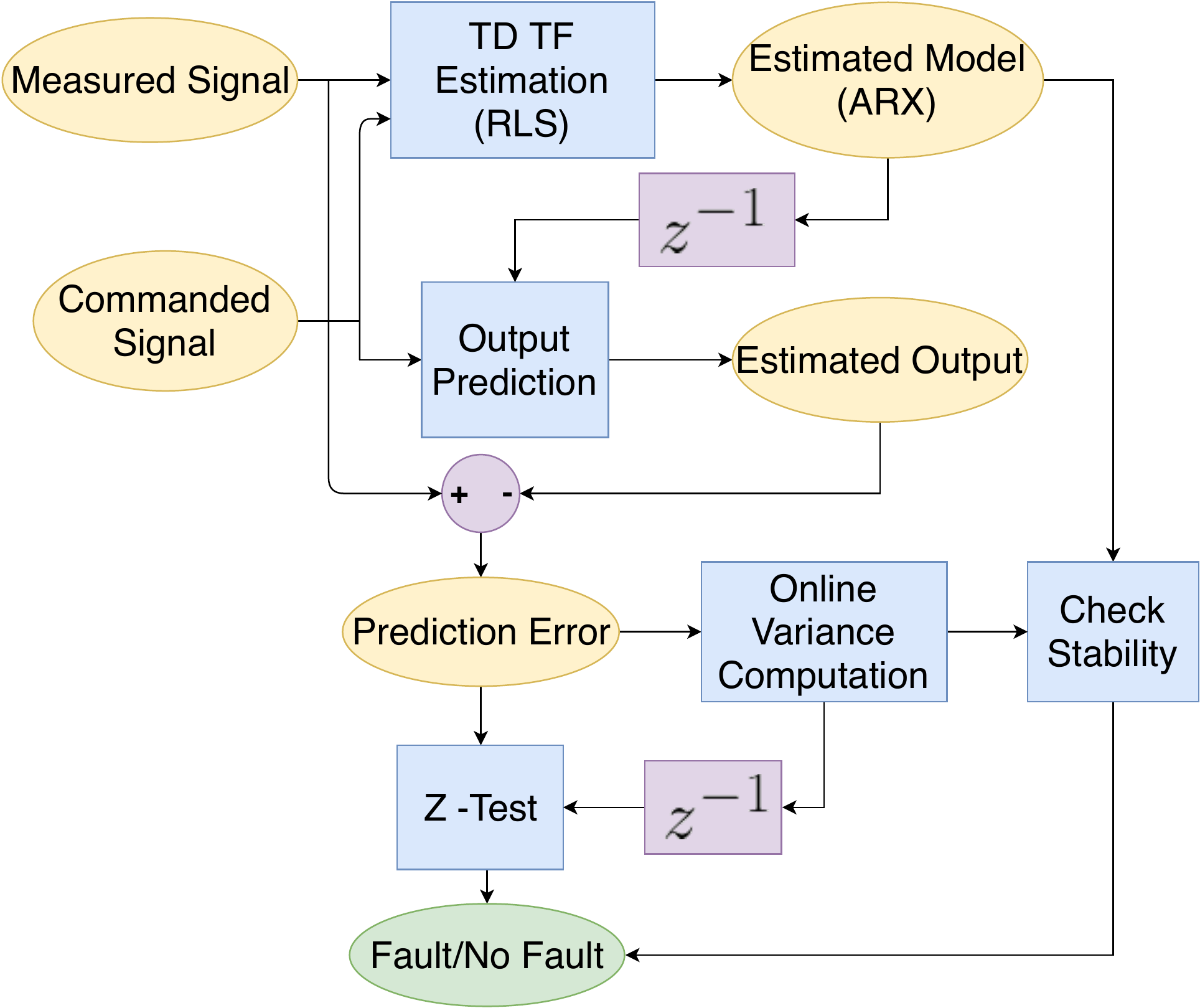}
    \caption{The flowchart of our method. The ARX Time-Domain Transfer Function is estimated for an input-output pair at each step using the RLS method. The error between the estimated output from the model and the measured output is used for both updating the model and for fault detection by calculating its Z-score.}
    \label{fig:flowchart}
\end{figure}

\section{APPROACH} \label{sec:method}

Our proposed method monitors a number of given input-output signal pairs of the aircraft. For each pair, the system identifies the relationship between the input and output signals and for each input signal sample, it predicts (estimates) the corresponding output sample. It solves the fault estimation problem by finding unexpected changes in the given output value compared to this predicted output sample. A high-level overview of the presented method is shown in Figure~\ref{fig:flowchart}.

This section discusses the chosen model for the relationship between the input and output signals, the algorithm used for the model estimation and the prediction of the output, the criteria used for anomaly detection, and the assumptions for our method.

\subsection{The Model} \label{sec:model}

The aircraft dynamics is highly nonlinear and cannot be captured using a linear model. However, in this approach, the goal is to model the relationship between two signals instead of the whole aircraft dynamics. Many signal pair are linearly related to each other; therefore we can use a linear model to estimate the relationship between such signals.

A general time-series linear model to capture the relationship (model) of the signals in a system takes the following form \cite{Ljung:1999:SIT:293154}:
\begin{equation}
y(k)=\frac{B(q^{-1})}{A(q^{-1})F(q^{-1})} u(k) + \frac{C(q^{-1})}{A(q^{-1})D(q^{-1})}n(k)
\end{equation}

\noindent where $q^{-1}$ is the \textit{lag operator} (also known as \textit{time-shift} operator), $A(q^{-1})$, $B(q^{-1})$, $C(q^{-1})$, $D(q^{-1})$, and $F(q^{-1})$ are polynomials of lag operators, $u(k)$ is the input signal and $y(k)$ is the output signal. White noise $n(k)$ is assumed to have zero mean value and constant variance. 

For our purposes, we chose an Autoregressive Exogenous (ARX) Time-Domain Transfer Function model, which is a specific case of the general linear model described above, where $C=D=F=1$. ARX model is defined as:
\begin{equation}
y(k)=\frac{B(q^{-1})}{A(q^{-1})}u(k) + \frac{1}{A(q^{-1})}n(k)
\end{equation}

\noindent which can be expanded as:
\begin{equation}
\begin{split}
&a_k y_k + a_{k-1} y_{k-1} + \hdots + a_{k-n_a} y_{k-n_a} \\
= \quad &b_k u_k + b_{k-1} u_{k-1} + \hdots + b_{k-n_b} u_{k-n_b} + n_k
\end{split}    
\end{equation}

The ARX model captures the input-output relationship using $n_b$ and $n_a$ coefficients for past input and output samples, respectively. In other words, to estimate (predict) the current output $y_k$, it is enough to have the coefficients $a_i$ and $b_j$, $n_a$ past outputs, $n_b$ past inputs and the current input $u_k$. As mentioned above, the white noise term $n(k)$ is assumed to have zero mean value and constant variance. 

\subsection{The Estimation Algorithm} \label{sec:rls-method}

Assuming a true unknown ARX model $\theta$ for the relationship of input-output signals, the estimation algorithm aims to calculate an estimated model $\hat\theta$ which converges to $\theta$ given enough samples. Recursive estimation methods aim to compute a new estimate $\hat\theta_{k}$ by a simple update to $\hat\theta_{k-1}$ when a new observation becomes available at iteration $k$. 

Our approach uses the Recursive Least Squares (RLS) method, which is an online optimization method (also famous as an adaptive filter algorithm) that recursively finds the coefficients to minimize a weighted linear least squares cost function related to the input signals \cite{hayes2009statistical}. This approach is in contrast to other algorithms such as Least Mean Squares (LMS) that aim to reduce the unweighted mean square error offline. In the derivation of RLS, the input signals are assumed to be deterministic, while for LMS and similar algorithms they are considered stochastic. Compared to most of the other methods, RLS exhibits a fast convergence \cite{1086206}. However, this benefit comes at the cost of higher computational complexity. The algorithm can be written in the following form:

\textit{Error term}, 
\begin{equation}
\hat e(k)=y(k) - \phi^T (k)\hat\theta (k-1)
\end{equation}

\textit{Gain matrix}, 
\begin{equation}
L(k)=\frac{C(k-1)\phi(k)}{1+\phi^T(k)C(k-1)\phi(k)}
\end{equation}

\textit{Estimated parameters}, 
\begin{equation}
\hat\theta (k)= \hat \theta(k-1) + L(k)\hat e(k)
\end{equation}

\textit{Covariance of estimated paramerters}, 
\begin{equation}
C(k)=C(k-1)-L(k)\phi^T(k)C(k-1)
\end{equation}

\noindent where, $\hat\theta(k-1)$ is a vector containing the stack of all the coefficients of the estimated ARX model before being updated with the new input; $\phi(k)$ is the vector containing all the new and past $y$ and $u$ signals stacked. $\phi^T (k)\hat\theta (k-1)$ is the prediction of the output using the new input and the model, which is used to calculate the error between the current output estimation and actual measured output given to the algorithm. This error term is used further for both updating the model itself and for fault detection (Sec.~\ref{sec:criteria}). To update the model, first, a gain matrix $L$ is calculated using the data vector $\phi(k)$ and the covariance matrix $C(k-1)$, then this gain is used to determine the magnitude of the update to the model $\hat\theta$. In the end, the covariance matrix is updated to be used for future model updates. The covariance matrix can be seen as an indicator for uncertainty in our model: a larger covariance means more uncertainty and results in a higher gain and more substantial updates to the model in each step; a smaller covariance means a less uncertainty in the model and tends to keep the model updates small. For initialization, the model can be set to a zero vector and the covariance matrix can be set to an identity matrix; however, to show a high uncertainty in the initial model and to increase the convergence rate, the covariance matrix can be set as a diagonal matrix with very large diagonal values.

After each step, the model is tested for stability. In practice, if $\|L(k)\hat e(k)\|_\infty < \epsilon$ for some time (with $\epsilon$ set to a small positive number), the model can be considered as stable for fault detection. 

\subsection{Criteria for Anomaly Detection} \label{sec:criteria}

Each step $k$ of the estimation algorithm described in Sec.~\ref{sec:rls-method} calculates an error term $\hat e(k)$ from the input $\phi$ and the estimated model $\hat\theta$. This error term captures the difference between the output $\hat y(k)$ predicted by the estimated model and the actual output $y(k)$ measured by the sensors. 
The variance of the error terms is calculated online using Welford's recursive method \cite{welford} shown below:
\begin{equation}
\begin{split}
&\Bar{x}_n = \Bar{x}_{n-1} + \frac{x_n - \Bar{x}_{n-1}}{n}\\
&M_{2,n} = M_{2, n-1} + (x_n - \Bar{x}_{n-1})(x_n - \Bar{x}_n)\\
&s^2_n = \frac{M_{2,n}}{n-1} \qquad, \qquad \sigma^2_n = \frac{M_{2,n}}{n}
\end{split}
\end{equation}

\begin{figure}[!t]
\centering
    \includegraphics[width=0.48\textwidth, height=0.32\textwidth]{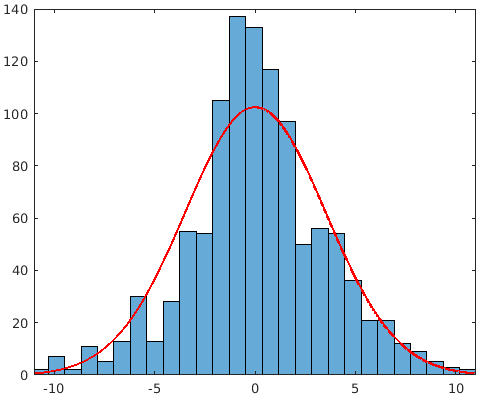}
    \vspace{-0.6cm} \caption{The distribution of roll estimation error from a sample sequence can be approximated by a zero-mean Gaussian distribution (the red curve).}
    \label{fig:distribution}
\end{figure}

Due to the Central Limit Theorem, the distribution for the error term $\hat e$ can be assumed as a Gaussian distribution. This can also be observed in Figure~\ref{fig:distribution} from the distribution of roll estimation errors from a sample flight sequence. Therefore, we can use the standard deviation for $\hat e$ to calculate the confidence of the new error term $\hat e(k)$. Assuming that the estimated model $\hat\theta$ can capture the relationship between the input-output pairs of the fault detection system, a high Z-score for $\hat e(k)$ shows that the output cannot be reliably predicted anymore, which may indicate an anomaly in the system. The variance of error terms $\hat e$ can be estimated from the start or from when the estimated model $\hat \theta$ is already stable (does not significantly change in each update) but new error terms with high Z-scores are flagged as anomaly only after variance is also stable after initial instability when the variance significantly changes after each update.

\subsection{Assumptions} \label{sec:assumptions}

The given approach estimates an ARX Discrete-Time model for each stream of input-output signals. This model assumes that the samples are available at a fixed sample rate. However, in practice, the method has shown some degree of robustness to small variations in the frequency of the stream of input-output samples. 

The input and output signals of the system can be chosen as any pair with approximately a linear relationship. Uncorrelated pairs are very unlikely to allow for the model estimation to stabilize, therefore adding uncorrelated pairs to the system will not help with the fault detection and will not result in false anomaly detections either. However, adding pairs with nonlinear relationships may result in stabilization of the estimated model and cause false detections later. The ideal choice for the input-output signals is the instantaneous commanded signal and its measured value for most of the signals. For example, the instantaneous roll commanded by the autopilot and the measured roll would be a useful pair for anomaly detection using this method.

Finally, the method assumes abrupt failures and may not be able to detect gradual faults. Besides, it is assumed that the failure happens after the initialization phase when the model and the variance prediction are already stable.

\section{TESTS AND RESULTS} \label{sec:tests}

\subsection{Hardware and Software} \label{sec:hardware-software}
\begin{figure}[!t]
\centering
    \includegraphics[width=0.48\textwidth]{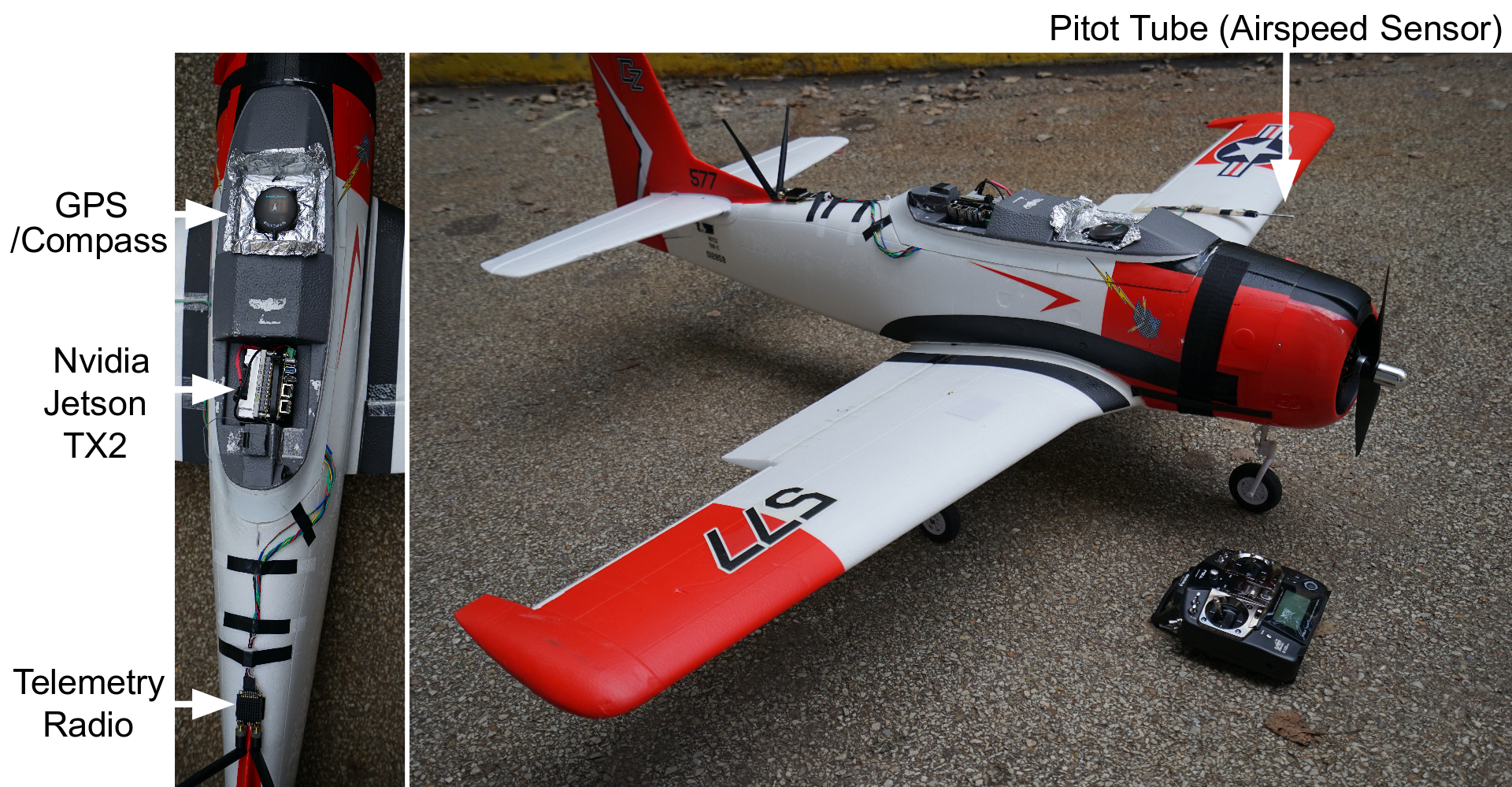}
    \vspace{-0.4cm} \caption{The Carbon-Z T-28 fixed-wing UAV platform equipped with an onboard computer and additional modules for our tests.}    
    \label{fig:platform}
\end{figure}

To test our proposed method, we implemented it in Linux Ubuntu 16.04 (Xenial) using C++'11 language and Robot Operating System (ROS) Kinetic Kame. The flight test platform is a custom modification of Carbon Z T-28, a fixed-wing UAV with 2 meters of wingspan and a central electric engine. Fig. \ref{fig:platform} shows the platform with the added sensors. In addition to a Pixhawk autopilot, a GPS module, a Pitot Tube airspeed sensor, and an Nvidia Jetson TX2 were added to the base platform. Pixhawk uses Ardupilot/ArduPlane v3.9.0 flight control software modified to publish the desired data for monitoring and to provide a direct way of imposing a failure on actuators in an autonomous flight. For safety purposes actuators were failing to work only until the flight mode was changed, giving the safety pilot ability to take control over the plane at any time that safety was going to be compromised. The trajectory controller for autonomous flights is a version of the controller used by Schopferer et al. \cite{azarakhsh-icuas18} to control the plane using an onboard computer vs. the original one that was controlling it from the ground station.

\subsection{Tests and Dataset} \label{sec:sub:tests}

The trial results were gathered to experimentally verify the capacity to recognize and detect different types of faults in AAV systems. We have tested eight types of failure consisting: Left/right/both ailerons stuck at zero position, elevator stuck at zero position, engine full power loss, rudder stuck at the left/right position, and both ailerons and rudder stuck at zero position. These failures are all actuator failures. We have provided the full dataset retrieved from our tests to facilitate future research in the area of automatic fault detection. It comprises full information containing ground truth (the time and type of the failure) from 22 flight tests on a fixed-wing AAV. The data from this work is further integrated into the AIR Lab Failure and Anomaly (ALFA) Dataset, which is presented in \cite{azarakhsh-ijrr19} and is available at \url{http://theairlab.org/alfa-dataset}.

We modified the Ardupilot firmware to publish commanded roll and pitch signals, then compute roll error and pitch error and use these four signals as input to our method. We defined roll/pitch errors as the difference between the commanded signal and the measured signal. For example, for roll error, the input pair would be $u_{roll}$ and $(y_{roll} - u_{roll})$. These input signals are generated with the frequency of about $25 Hz$, and we assume a Gaussian distribution for noise. It can be mathematically proven that for unlimited number of signal samples, both roll/pitch and roll/pitch errors can capture the exact same amount of statistical information; however, on account of the limited number of samples being fed to the system, roll/pitch error could sometimes capture better model and hence result in better performance. We use Z-score of 4.5, which corresponds to the 99.99931\% confidence interval. It takes at least 8 seconds (200 samples) for the model to become stable. In our tests, we used the inputs from the last second (25 samples) for the estimation of the new output.

\subsection{Results} \label{sec:sub:results}

\begin{table}[!t]
\centering
\caption{Test Statistics}
\label{table:stats}
\resizebox{8.6cm}{!}{%
\begin{tabular}{|
>{\columncolor[HTML]{C0C0C0}}c |c|c|c|c|c|}
\hline
\cellcolor[HTML]{CBCEFB}\textbf{\begin{tabular}[c]{@{}c@{}}Failure\\ Type\end{tabular}} & \cellcolor[HTML]{CBCEFB}\textbf{\begin{tabular}[c]{@{}c@{}}\# of \\ tests\end{tabular}} & \cellcolor[HTML]{CBCEFB}\textbf{\begin{tabular}[c]{@{}c@{}}Flight\\ Time(s)\end{tabular}} & \cellcolor[HTML]{CBCEFB}\textbf{\begin{tabular}[c]{@{}c@{}}Avg.\\Detection\\Time(s)\end{tabular}} & \cellcolor[HTML]{CBCEFB}\textbf{\begin{tabular}[c]{@{}c@{}}Max\\Detection\\Time(s)\end{tabular}} & \cellcolor[HTML]{CBCEFB}\textbf{\begin{tabular}[c]{@{}c@{}}Accuracy\\(\%)\end{tabular}} \\ \hline
Engine & 7 & 665 & 2.28 & 3.37 & 100 \\ \hline
Rudder & 3 & 171 & 0.21 & 0.25 & -\\ \hline
Elevator & 2 & 181 & 0.36 & 0.36 & -\\ \hline
Aileron & 4 & 340 & 3.31 & 5.6 & -\\ \hline 
Rudder/Aileron & 1 & 116 & 3.48 & 3.48 & -\\ \hline 
No Failure & 5 & 262 & - & - & -\\ \hline \hline
\textbf{Total} & 22 & 1735 & 2.02 & 5.6 & 86.36\\ \hline
\end{tabular}
}
\end{table}

Table. \ref{table:stats} presents the statistics over 22 flight tests with different types of failure. We evaluate our method by utilizing different performance metrics. One performance metric is the number of false detections (False Positives and False Negatives). Our test results indicate a total number of 2 False Positive (FP) and 2 False Negative (FN) detections out of 22 tests, where 1 FP and 1 FN happen in a single flight (it announces anomaly but before the failure actually happens). With the 19 correct sequences, our method results in $86.36\%$ accuracy, $88.23\%$ precision and $88.23\%$ recall (sensitivity) over 22 flight tests. The evaluation metrics used for our calculations are explained in more details in \cite{azarakhsh-ijrr19}.

\begin{figure}[!t]
\centering
    \includegraphics[width=0.48\textwidth]{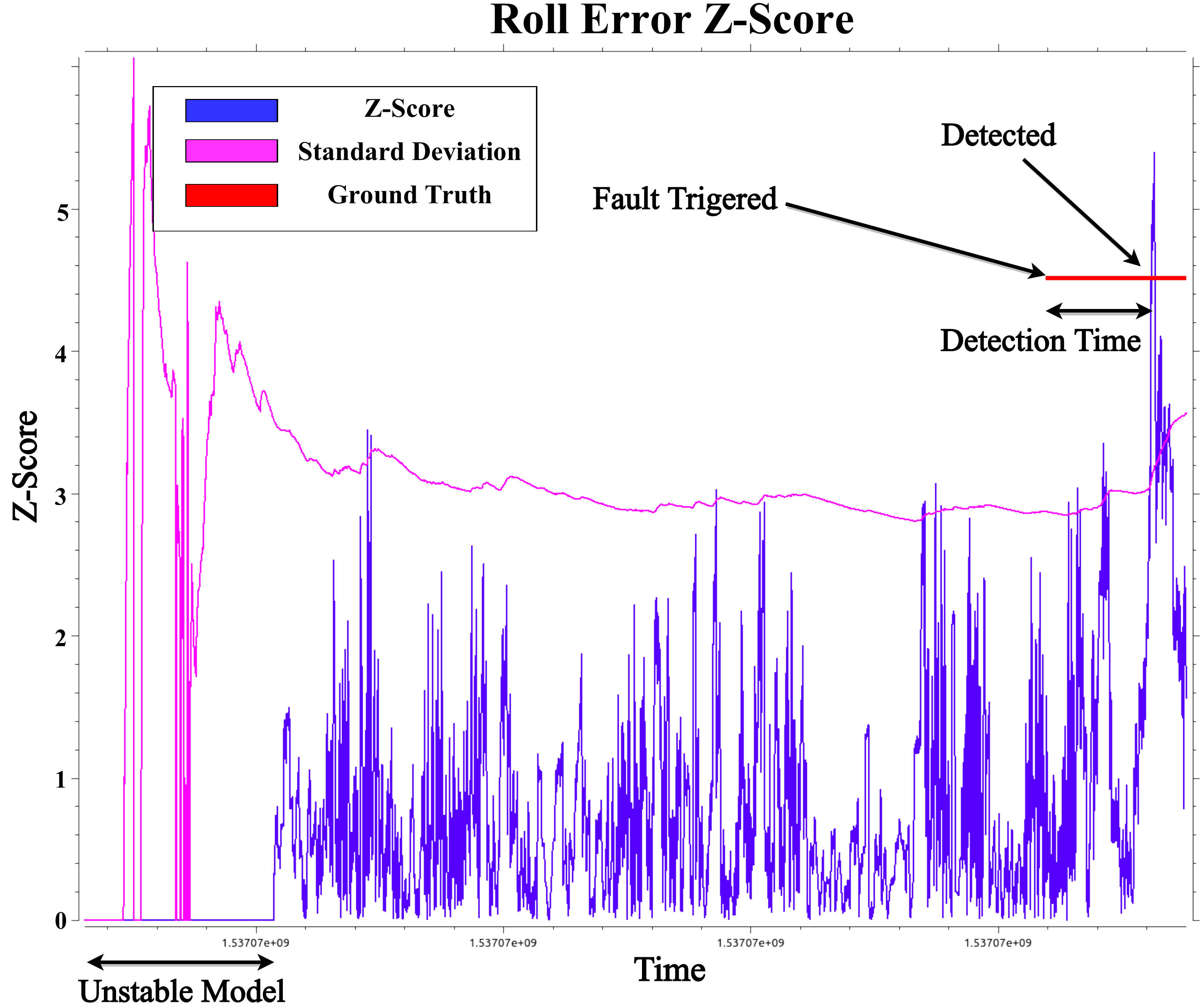}
    \caption{Z-Score vs Roll Error input for an engine failure flight test. Fault is triggered for this input signal.}    
    \label{fig:plot1}
\end{figure}

\begin{figure}[!t]
\centering
    \includegraphics[width=0.48\textwidth]{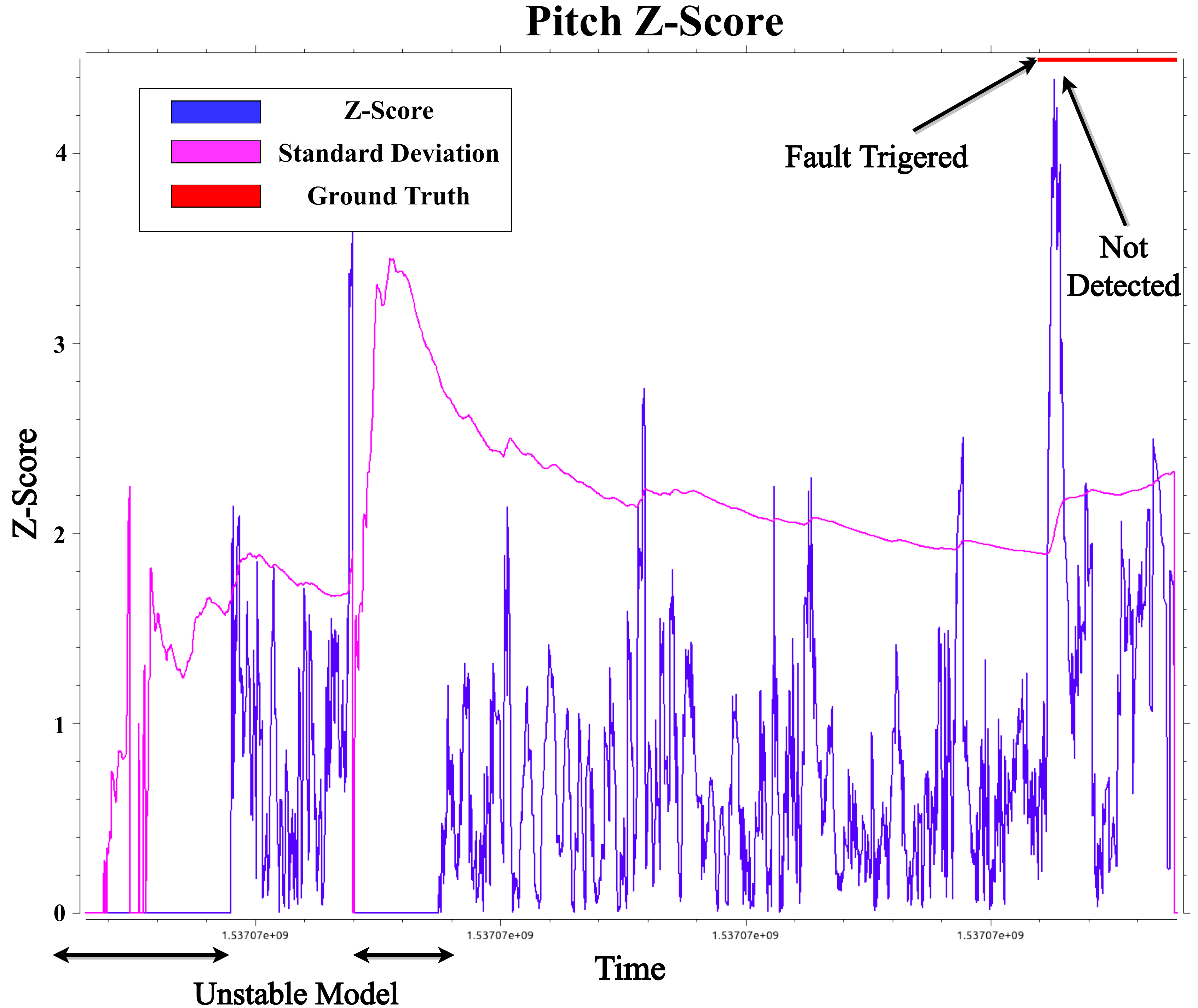}
    \caption{Z-Score vs Pitch input for an engine failure flight test. Fault is not triggered for this input signal. However, this failure is already detected with two other input signals.}    
    \label{fig:plot3}
\end{figure}

Figure~\ref{fig:plot1} and shows how the system finds an anomaly from Roll Error commanded/measured pairs when an engine failure happens. After the initial stabilization phase, the Z-score of the prediction error tends to be significantly less than the set threshold of 4.5 for the anomaly. Figure~\ref{fig:plot3} shows the system monitoring Pitch commanded/measured pair not being able to detect the failure at the same time despite having a spike in the prediction error Z-score when the failure happens. In this particular example, the Roll Error signals are the one announcing the anomaly in the system. The figures also show how the variance of the prediction error stabilizes after the model stabilization. 

While this work and \cite{azarakhsh-ijrr19} provide a dataset suitable for benchmarking different methods, to the best of our knowledge, there has been no benchmark dataset available prior to this work to enable direct comparison with the published results of similar works like Venkataraman et al. \cite{VENKATARAMAN2019365} and Bauer et al. \cite{BAUER2018600}. The use of our dataset in the future will enable the comparison of methods to the state-of-the-art.

\section{DISCUSSION AND FUTURE WORK} \label{sec:discussion}

The proposed method provides a simple way to add a Fault Detection System to an autonomous plane. The method works by monitoring pairs of input-output signals for unexpected output values for the provided input signals. The selected pairs should have a linear relationship with each other in order to be useful in the fault detection process; therefore they need to be selected carefully. To simplify this manual process, another system can automatically find correlated input-output pairs (similar to the work done by Khalastchi et al. \cite{p115-khalastchi}) and feed them to the current system for monitoring. 

The current system creates a separate single-input single-output model for each input-output signal pair. In many cases, output signals get influenced by two or more signals. One idea to handle these relationships can be to find these input-output signal sets from the model of the aircraft and perform a multidimensional model estimation for fault detection. However, this approach will make the current model-independent method dependent on the specific aircraft type. To avoid this dependency, it is possible to add all inputs and all outputs to a single multidimensional model. This will be at a much higher computational cost, and the model may face convergence issues.

Most of the published results so far, either have not reported any results on real flights or have not clearly explained their exact way of testing. By providing our dataset, we hope that it will save other researchers data collection time and help them in assessing their implemented methods. We have integrated this dataset into a larger ALFA dataset \cite{azarakhsh-ijrr19}, and we aim to further provide mid-air sensor and actuator failures for different types of AAVs.

The lack of excitation in the signals may result in poor model estimations in RLS-based methods. A possible future improvement can be turning the model estimation off when there is not enough excitation \cite{Improved-least}. 

Finally, to capture nonlinear relationships between input-output signal pairs, it is suggested to try using a Nonlinear Autoregressive Exogenous (NARX) model with a nonlinear model estimation in future developments. 

\section{CONCLUSIONS} \label{sec:conclusions}

In this paper, we proposed a method for real-time fault detection using the RLS algorithm. We implemented our method on an autonomous fixed-wing aircraft, and our experiments show a precision of $88.23\%$, recall of $88.23\%$ and $86.36\%$ accuracy during 22 flight tests. Our results show improvement of both accuracy and detection time for fault detection in AAVs compared to the other available approaches. The key contributions that distinguish our work from the previous research in this area are that our method estimates the model between correlated input-output signal pair online and therefore does not depend on a specific model, can detect a wide range of anomalies, and is easy to integrate to any system. Furthermore, we have provided our dataset with 22 tests including eight different types for mid-air failures for public use.

\addtolength{\textheight}{-3.2cm}   



\section*{ACKNOWLEDGMENT}

This project became possible due to the support of Near Earth Autonomy (NEA). Also, the authors would like to thank Mark DeLouis for his expertise as a pilot and his help with the platform and hardware issues during the project.

\bibliographystyle{IEEEtran}

\begin{thebibliography}{10}
\providecommand{\url}[1]{#1}
\csname url@samestyle\endcsname
\providecommand{\newblock}{\relax}
\providecommand{\bibinfo}[2]{#2}
\providecommand{\BIBentrySTDinterwordspacing}{\spaceskip=0pt\relax}
\providecommand{\BIBentryALTinterwordstretchfactor}{4}
\providecommand{\BIBentryALTinterwordspacing}{\spaceskip=\fontdimen2\font plus
\BIBentryALTinterwordstretchfactor\fontdimen3\font minus
  \fontdimen4\font\relax}
\providecommand{\BIBforeignlanguage}[2]{{%
\expandafter\ifx\csname l@#1\endcsname\relax
\typeout{** WARNING: IEEEtran.bst: No hyphenation pattern has been}%
\typeout{** loaded for the language `#1'. Using the pattern for}%
\typeout{** the default language instead.}%
\else
\language=\csname l@#1\endcsname
\fi
#2}}
\providecommand{\BIBdecl}{\relax}
\BIBdecl

\bibitem{faa}
\BIBentryALTinterwordspacing
F.~A.~A. (FAA). (2004) Title 14 code of federal regulations part 145 approved
  training program. [Online]. Available:
  \url{https://www.faa.gov/aircraft/air\_cert/production\_approvals/14cfr\_amen
  dments/}
\BIBentrySTDinterwordspacing

\bibitem{8014173}
M.~Kuric, B.~Lacevic, N.~Osmic, and A.~Tahirovic, ``Rls-based fault-tolerant
  tracking control of multirotor aerial vehicles,'' in \emph{2017 IEEE
  International Conference on Advanced Intelligent Mechatronics (AIM)}, July
  2017, pp. 1148--1153.

\bibitem{954410017691794}
\BIBentryALTinterwordspacing
W.~Han, Z.~Wang, and Y.~Shen, ``Fault estimation for a quadrotor unmanned
  aerial vehicle by integrating the parity space approach with recursive least
  squares,'' \emph{Proceedings of the Institution of Mechanical Engineers, Part
  G: Journal of Aerospace Engineering}, vol. 232, no.~4, pp. 783--796, 2018.
  [Online]. Available: \url{https://doi.org/10.1177/0954410017691794}
\BIBentrySTDinterwordspacing

\bibitem{6564801}
X.~Qi, D.~Theilliol, J.~Qi, Y.~Zhang, and J.~Han, ``A literature review on
  fault diagnosis methods for manned and unmanned helicopters,'' in \emph{2013
  International Conference on Unmanned Aircraft Systems (ICUAS)}, May 2013, pp.
  1114--1118.

\bibitem{AReviewonFault}
\BIBentryALTinterwordspacing
X.~Qi, J.~Qi, D.~Theilliol, Y.~Zhang, J.~Han, D.~Song, and C.~Hua, ``A review
  on fault diagnosis and fault tolerant control methods for single-rotor aerial
  vehicles,'' \emph{Journal of Intelligent {\&} Robotic Systems}, vol.~73,
  no.~1, pp. 535--555, Jan 2014. [Online]. Available:
  \url{https://doi.org/10.1007/s10846-013-9954-z}
\BIBentrySTDinterwordspacing

\bibitem{1-s2.0-S0967066100000460-main}
\BIBentryALTinterwordspacing
J.~Melody, T.~Başar, W.~Perkins, and P.~Voulgaris, ``Parameter identification
  for inflight detection and characterization of aircraft icing,''
  \emph{Control Engineering Practice}, vol.~8, no.~9, pp. 985--1001, 2000.
  [Online]. Available:
  \url{http://www.sciencedirect.com/science/article/pii/S0967066100000460}
\BIBentrySTDinterwordspacing

\bibitem{7526603}
A.~Ansari and D.~S. Bernstein, ``Aircraft sensor fault detection using state
  and input estimation,'' in \emph{2016 American Control Conference (ACC)},
  July 2016, pp. 5951--5956.

\bibitem{VANSCHRICK1997959}
\BIBentryALTinterwordspacing
D.~van Schrick, ``Remarks on terminology in the field of supervision, fault
  detection and diagnosis,'' \emph{IFAC Proceedings Volumes}, vol.~30, no.~18,
  pp. 959--964, 1997, iFAC Symposium on Fault Detection, Supervision and Safety
  for Technical Processes (SAFEPROCESS 97), Kingston upon Hull, UK, 26-28
  August 1997. [Online]. Available:
  \url{http://www.sciencedirect.com/science/article/pii/S1474667017425249}
\BIBentrySTDinterwordspacing

\bibitem{7069265}
Z.~Gao, C.~Cecati, and S.~X. Ding, ``A survey of fault diagnosis and
  fault-tolerant techniques—part i: Fault diagnosis with model-based and
  signal-based approaches,'' \emph{IEEE Transactions on Industrial
  Electronics}, vol.~62, no.~6, pp. 3757--3767, June 2015.

\bibitem{icing1}
\BIBentryALTinterwordspacing
A.~Cristofaro, T.~A. Johansen, and A.~P. Aguiar, ``Icing detection and
  identification for unmanned aerial vehicles using adaptive nested multiple
  models,'' \emph{International Journal of Adaptive Control and Signal
  Processing}, vol.~31, no.~11, pp. 1584--1607, 2017. [Online]. Available:
  \url{https://onlinelibrary.wiley.com/doi/abs/10.1002/acs.2787}
\BIBentrySTDinterwordspacing

\bibitem{10.10072F978-90-481-9707-143}
\BIBentryALTinterwordspacing
G.~Ducard, \emph{Actuator Fault Detection in UAVs}.\hskip 1em plus 0.5em minus
  0.4em\relax Dordrecht: Springer Netherlands, 2015, pp. 1071--1122. [Online].
  Available: \url{https://doi.org/10.1007/978-90-481-9707-1_43}
\BIBentrySTDinterwordspacing

\bibitem{Khalastchi:2018:FDD:3177787.3146389}
\BIBentryALTinterwordspacing
E.~Khalastchi and M.~Kalech, ``On fault detection and diagnosis in robotic
  systems,'' \emph{ACM Computing Surveys (CSUR)}, vol.~51, no.~1, pp.
  9:1--9:24, Jan. 2018. [Online]. Available:
  \url{http://doi.acm.org/10.1145/3146389}
\BIBentrySTDinterwordspacing

\bibitem{UAVSecurity0167FrDTT2-04}
Z.~Birnbaum, A.~Dolgikh, V.~Skormin, E.~O'Brien, D.~Muller, and
  C.~Stracquodaine, ``Unmanned aerial vehicle security using behavioral
  profiling,'' in \emph{2015 International Conference on Unmanned Aircraft
  Systems (ICUAS)}, June 2015, pp. 1310--1319.

\bibitem{birnbaum2015}
\BIBentryALTinterwordspacing
Z.~Birnbaum, A.~Dolgikh, V.~Skormin, E.~O'Brien, and D.~Muller, ``Unmanned
  aerial vehicle security using recursive parameter estimation,'' \emph{Journal
  of Intelligent {\&} Robotic Systems}, vol.~84, no.~1, pp. 107--120, Dec 2016.
  [Online]. Available: \url{https://doi.org/10.1007/s10846-015-0284-1}
\BIBentrySTDinterwordspacing

\bibitem{p115-khalastchi}
\BIBentryALTinterwordspacing
E.~Khalastchi, G.~A. Kaminka, M.~Kalech, and R.~Lin, ``Online anomaly detection
  in unmanned vehicles,'' in \emph{The 10th International Conference on
  Autonomous Agents and Multiagent Systems - Volume 1}, ser. AAMAS '11.\hskip
  1em plus 0.5em minus 0.4em\relax Richland, SC: International Foundation for
  Autonomous Agents and Multiagent Systems, 2011, pp. 115--122. [Online].
  Available: \url{http://dl.acm.org/citation.cfm?id=2030470.2030487}
\BIBentrySTDinterwordspacing

\bibitem{10873863}
\BIBentryALTinterwordspacing
L.~R. Cork, R.~A. Walker, and S.~Dunn, ``Fault detection, identification and
  accommodation techniques for unmanned airborne vehicles,'' in
  \emph{Australian International Aerospace Congress}.\hskip 1em plus 0.5em
  minus 0.4em\relax Melbourne: Australian International Aerospace Congress
  (AIAC), 2005, pp. 230--235. [Online]. Available:
  \url{https://eprints.qut.edu.au/1729/}
\BIBentrySTDinterwordspacing

\bibitem{neural-observer}
\BIBentryALTinterwordspacing
A.~Abbaspour, P.~Aboutalebi, K.~K. Yen, and A.~Sargolzaei, ``Neural adaptive
  observer-based sensor and actuator fault detection in nonlinear systems:
  Application in uav,'' \emph{ISA Transactions}, vol.~67, pp. 317--329, 2017.
  [Online]. Available:
  \url{http://www.sciencedirect.com/science/article/pii/S0019057816306656}
\BIBentrySTDinterwordspacing

\bibitem{sensors-17-02243}
\BIBentryALTinterwordspacing
R.~Sun, Q.~Cheng, G.~Wang, and W.~Y. Ochieng, ``A novel online data-driven
  algorithm for detecting uav navigation sensor faults,'' \emph{Sensors},
  vol.~17, no.~10, 2017. [Online]. Available:
  \url{http://www.mdpi.com/1424-8220/17/10/2243}
\BIBentrySTDinterwordspacing

\bibitem{4252507}
\BIBentryALTinterwordspacing
L.~{Cork} and R.~{Walker}, ``Sensor fault detection for uavs using a nonlinear
  dynamic model and the imm-ukf algorithm,'' in \emph{2007 Information,
  Decision and Control}, Feb 2007, pp. 230--235. [Online]. Available:
  \url{https://ieeexplore.ieee.org/document/4252507}
\BIBentrySTDinterwordspacing

\bibitem{4603358}
\BIBentryALTinterwordspacing
J.~{Qi} and J.~{Han}, ``Fault adaptive control for ruav actuator failure with
  unscented kalman filter,'' in \emph{2008 3rd International Conference on
  Innovative Computing Information and Control}, June 2008, pp. 169--169.
  [Online]. Available: \url{https://ieeexplore.ieee.org/document/4603358}
\BIBentrySTDinterwordspacing

\bibitem{IZADI20116343}
\BIBentryALTinterwordspacing
H.~A. Izadi, Y.~Zhang, and B.~W. Gordon, ``Fault tolerant model predictive
  control of quad-rotor helicopters with actuator fault estimation,''
  \emph{IFAC Proceedings Volumes}, vol.~44, no.~1, pp. 6343 -- 6348, 2011, 18th
  IFAC World Congress. [Online]. Available:
  \url{http://www.sciencedirect.com/science/article/pii/S1474667016446227}
\BIBentrySTDinterwordspacing

\bibitem{Ljung:1999:SIT:293154}
L.~Ljung, Ed., \emph{System Identification (2Nd Ed.): Theory for the
  User}.\hskip 1em plus 0.5em minus 0.4em\relax Upper Saddle River, NJ, USA:
  Prentice Hall PTR, 1999.

\bibitem{hayes2009statistical}
M.~H. Hayes, \emph{Statistical digital signal processing and modeling}.\hskip
  1em plus 0.5em minus 0.4em\relax John Wiley \& Sons, 2009.

\bibitem{1086206}
E.~Eweda and O.~Macchi, ``Convergence of the rls and lms adaptive filters,''
  \emph{IEEE Transactions on Circuits and Systems}, vol.~34, no.~7, pp.
  799--803, July 1987.

\bibitem{welford}
\BIBentryALTinterwordspacing
B.~P. Welford, ``Note on a method for calculating corrected sums of squares and
  products,'' \emph{Technometrics}, vol.~4, no.~3, pp. 419--420, 1962.
  [Online]. Available: \url{http://www.jstor.org/stable/1266577}
\BIBentrySTDinterwordspacing

\bibitem{azarakhsh-icuas18}
S.~Schopferer, J.~S. Lorenz, A.~Keipour, and S.~Scherer, ``Path planning for
  unmanned fixed-wing aircraft in uncertain wind conditions using trochoids,''
  in \emph{2018 International Conference on Unmanned Aircraft Systems (ICUAS)},
  June 2018, pp. 503--512.

\bibitem{azarakhsh-ijrr19}
\BIBentryALTinterwordspacing
A.~Keipour, M.~Mousaei, and S.~Scherer, ``Alfa: A dataset for uav fault and
  anomaly detection,'' \emph{The International Journal of Robotics Research},
  vol.~0, no.~0, pp. 1--6, oct 2020. [Online]. Available:
  \url{https://doi.org/10.1177/0278364920966642}
\BIBentrySTDinterwordspacing

\bibitem{VENKATARAMAN2019365}
\BIBentryALTinterwordspacing
R.~Venkataraman, P.~Bauer, P.~Seiler, and B.~Vanek, ``Comparison of fault
  detection and isolation methods for a small unmanned aircraft,''
  \emph{Control Engineering Practice}, vol.~84, pp. 365--376, 2019. [Online].
  Available:
  \url{http://www.sciencedirect.com/science/article/pii/S0967066118303708}
\BIBentrySTDinterwordspacing

\bibitem{BAUER2018600}
\BIBentryALTinterwordspacing
P.~Bauer, R.~Venkataraman, B.~Vanek, P.~J. Seiler, and J.~Bokor, ``Fault
  detection and basic in-flight reconfiguration of a small uav equipped with
  elevons,'' \emph{IFAC-PapersOnLine}, vol.~51, no.~24, pp. 600--607, 2018,
  10th IFAC Symposium on Fault Detection, Supervision and Safety for Technical
  Processes SAFEPROCESS 2018. [Online]. Available:
  \url{http://www.sciencedirect.com/science/article/pii/S2405896318323498}
\BIBentrySTDinterwordspacing

\bibitem{Improved-least}
\BIBentryALTinterwordspacing
N.~R. Sripada and D.~G. Fisher, ``Improved least squares identification,''
  \emph{International Journal of Control}, vol.~46, no.~6, pp. 1889--1913,
  1987. [Online]. Available: \url{https://doi.org/10.1080/00207178708934023}
\BIBentrySTDinterwordspacing

\end{thebibliography}

\end{document}